\documentclass[prb,showpacs,preprintnumbers,amsmath,amssymb,printfigures,twocolumn,superscriptaddress]{revtex4-1}
\usepackage[colorlinks]{hyperref}
\usepackage{graphicx}
\usepackage{dcolumn}
\usepackage{amsmath}
\usepackage{lipsum}
\usepackage{amssymb}
\usepackage{amsthm}
\usepackage{mathtools}
\usepackage{bm}

\begin{document}
\title{Piezoelectricity in two-dimensional materials: a comparative study between lattice dynamics and ab-initio calculations }

\author{K. H. Michel}
\email{ktdm@skynet.be}
\affiliation{Departement Fysica,
Universiteit Antwerpen, Groenenborgerlaan 171, B-2020 Antwerpen, Belgium.}
\author{ D. \c{C}ak{\i}r}
\email{deniz.cakir@und.edu}
\affiliation{Departement Fysica,
Universiteit Antwerpen, Groenenborgerlaan 171, B-2020 Antwerpen, Belgium.}
\affiliation{Department of Physics and Astrophysics, University of North Dakota, Grand Forks, ND, 58202, USA}
\author{C. Sevik}
\email{csevik@anadolu.edu.tr}
\affiliation{Department of Mechanical Engineering, Faculty of Engineering, Anadolu University, Eskisehir, TR 26555, Turkey.}
\author{F. M. Peeters}
\email{francois.peeters@uantwerpen.be}
\affiliation{Departement Fysica,
Universiteit Antwerpen, Groenenborgerlaan 171, B-2020 Antwerpen, Belgium.}

\date{\today}

\begin{abstract}
Elastic constant C$_{11}$ and piezoelectric stress constant $e_{1,11}$ of two-dimensional (2D) dielectric materials comprising  h-BN, 2H MoS$_2$
and other transition metal dichalcogenides (TMDCs) and -dioxides (TMDOs) are calculated using lattice dynamical theory. The results are compared with corresponding quantities obtained by $ab$-$initio$ calculations.  We identify the difference between clamped-ion and relaxed-ion contributions with the dependence on inner strains which are due to the relative displacements of the ions in the unit cell. Lattice dynamics allows to express the inner strains contributions in terms of microscopic quantities such as effective ionic charges and opto-acoustical couplings, which allows us to clarify  differences in the piezoelectric behavior between h-BN versus MoS$_2$. Trends in the different microscopic quantities as functions of   atomic composition are discussed. 
\end{abstract}

\pacs{63.22.-m,77.65.-j}

\maketitle

\section{Introduction}
%
%
Piezoelectricity is the manifestation of electro-mechanical coupling that is present in non-centrosymmetric  dielectric crystals\cite{1}. An electric polarization
occurs in response to macroscopic strains, its converse is the change of shape of the crystal upon application of an electric field. Numerous technological
applications are based on the use of bulk (three-dimensional) crystals and ceramics. Piezoelectricity in two-dimensional and in layered crystals is still at the stage of fundamental research. 

The synthesis of hexagonal boron-nitride (h-BN) nanotubes\cite{2,3} has stimulated theoretical work on polarization and piezoelectricity in BN nanotubes and its dependence on their topology\cite{4,5,6,nw1}. Even more the discovery of graphene and other 2D crystals like h-BN and MoS$_2$ has opened the road for the study of mechanical and electric properties of a new class of materials with controlled composition and number of atomic layers (i.e. its thickness)\cite{7,8}.

2D h-BN, the structurally most simple dielectric crystal, has been used by as a prototype in studies of piezo- and flexoelectricity by ab-initio calculations\cite{6,9} . Lattice dynamical theory has been used to study piezoelectricity in single and multilayered crystals  with application to h-BN as a specific example\cite{10,11}. 
 
First-principles calculations on 2H-MoS$_2$ and other transition metal dichalcogenides (TMDCs) have revealed that piezoelectricity in these materials is even more
pronounced than in 2D h-BN\cite{12}.   Notice that h-BN and MoS$_2$ have the same layer number $N$ dependent symmetry, D$_{3h}$ and D$_{3d}$, for odd and even $N$, respectively\cite{13} . Piezoelectricity has been measured recently in MoS$_2$ layered crystals\cite{14,15}. In agreement with predictions originally made by analytical theory\cite{11} it was found that only crystals with an uneven number of layers are piezoelectric with strength  of the piezoelectric stress coefficient decreasing as 1/$N$ with increasing $N$. Piezoelectric and elastic properties of a broad range of transition metal (group IV B, VI B) dichalcogenides and dioxides monolayers have been investigated by means of first-principles calculations. It was found that Ti-, Zn-, Sn-, and Cr-based TMDCs and TMDOs have much better piezoelectric properties as compared to Mo- and W-based materials\cite{16}. 
 
 The analytical microscopic theory of piezoelectricity  goes back to Born and is based on the same lattice dynamical concepts as the theory of elastic properties in ionic crystals\cite{17,19}. On the other hand, nowadays first-principles density functional investigations are the method of choice for studying piezoelectric and elastic properties of solids\cite{abinitio,abinitio2,abinitio3}. The aim of the present paper is to confront  the results of analytical  theory with first-principles calculations in 2D materials. Such an investigation will allow us to elucidate trends in the evolution of physical properties
as a function of ionic composition in classes of similar materials with the same crystal structure such as TMDCs and TMDOs. 

The content of the paper is as follows. In Sec.~\ref{sect2} we recall basic concepts of the analytical theory of elastic constants and piezoelectric moduli in 2D crystals. Next (Sect.~\ref{sect3}) we apply the analytical theory to 2D crystals with D$_{3h}$ point group symmetry, treating h-BN and 2H-MoS$_2$ as specific examples. In Sect.~\ref{sect4} we will give a discussion of numerical results obtained from $ab$-initio calculations and analytical theory for a broad range of 2D TMDCs and TMDOs materials. 

\section{Basic theory}\label{sect2}
\subsection{Elastic constants}
The analytical theory of elasticity of crystals of composite structures is based on Born's method of long waves \cite{17}.
In three-dimensional (3D) ionic crystals the formulation is complicated  by the appearance of  divergent results due to 
the long  range Coulomb forces. In 2D crystals these divergences are absent \cite{18} and we can restrict ourselves to the simple
treatment. In particular, the corrections to the elastic constants due to piezoelectricity, well known in 3D crystals\cite{landau}, vanish in the 2D case \cite{10}. The basic theoretical quantity in lattice dynamics is the dynamical matrix $\mathrm{D}(\overrightarrow{q})$, here $\overrightarrow{q}$=($q_x$,$q_y$) is a wave vector in the 2D Brillouin zone (BZ). One expands the dynamical matrix $\mathrm{D}(\overrightarrow{q})$ in powers of components of the wave vector $\overrightarrow{q}$. The dynamical
matrix of order 3$s$ has the elements $\mathrm{D}_{ij}^{\kappa \kappa'}(\overrightarrow{q})$, where $i$ ($j$) = $x, y, z$ labels the cartesian displacements and $\kappa$=1, 2,...$s$ refer to the particles in the unit cell. The long wave expansion for non-primitive crystals reads

\begin{equation}\label{1}
\mathrm{D}_{ij}^{\kappa \kappa'}(\overrightarrow{q})=\mathrm{D}_{ij}^{\kappa \kappa'(0)}+i\sum_k \mathrm{D}_{ij,k}^{\kappa \kappa'(1)}q_k+\frac{1}{2} \sum_{kl}\mathrm{D}_{ij,kl}^{\kappa \kappa'(2)}q_kq_l+... .
\end{equation}

The matrices $\mathrm{D}^{(0)}$, $\mathrm{D}^{(1)}$ and $\mathrm{D}^{(2)}$ represent the moments of the atomic force constants \cite{17,19} and have the following meaning:
$\mathrm{D}^{(0)}$  determines  the optical phonon frequencies at the $\Gamma$ point of the BZ,  $\mathrm{D}^{(1)}$ accounts for the coupling 
between optical and strains, $\mathrm{D}^{(2)}$ accounts for the coupling between strains. Notice that the elements $\mathrm{D}_{ij,k}^{\kappa \kappa' (1)}(\overrightarrow{q})$ are only different from zero if the ions are not centers of symmetry, and that $\mathrm{D}_{ij,k}^{\kappa \kappa' (1)}(\overrightarrow{q})$=-$\mathrm{D}_{ij,k}^{\kappa' \kappa (1)}(\overrightarrow{q})$. By perturbation theory one eliminates the optical displacements and obtains the acoustic dynamical matrix

\begin{equation}\label{2}
\mathrm{\hat{D}}_{ij}(\overrightarrow{q})=\frac{1}{\rho_{2D}}\sum_{kl}\{[ij,kl]+(ik,jl)\} q_kq_l.
\end{equation}
Here $\rho_{2D}$ is the surface mass density, $[ij,kl]$  stands for

\begin{equation}\label{3a}
[ij,kl]=\frac{1}{2A_{2D}}\sum_{\kappa \kappa'} (m_{\kappa}m_{\kappa'})^{1/2}\mathrm{D}_{ij,kl}^{\kappa \kappa'(2)},
\end{equation}
and 

\begin{widetext}
\begin{equation}\label{3b}
(ik,jl)=-\frac{1}{2A_{2D}}\sum_{\kappa \kappa'} \sum_{hp}\Gamma^{\kappa \kappa'}_{hp}[\sum_{\kappa''}(m_{\kappa''})^{1/2} \mathrm{D}_{hi,k}^{\kappa \kappa''(1)}] 
[\sum_{\kappa'''}(m_{\kappa'''})^{1/2} \mathrm{D}_{pj,k}^{\kappa' \kappa'''(1)}].
\end{equation}
\end{widetext}
Here $A_{2D}$ is the area of the unit cell, $m_{\kappa}$ is the mass of particle $\kappa$ and $i, j, k, l$ are Cartesian indices. The quantity  $\Gamma_{hp}^{\kappa \kappa'}$ is given by

\begin{equation}\label{4}
\Gamma_{hp}^{\kappa \kappa'}=\sum_{\lambda}\frac{\xi^{(\lambda)}(\kappa h)\xi^{(\lambda)}(\kappa' p)}{\omega^2_{\lambda}},
\end{equation}
where $\omega_{\lambda}$  and $\xi^{(\lambda)}$ are the optical eigenfrequencies and eigenvectors of the matrix $\mathrm{D}^{(0)}$, respectively. As is obvious
from Eqs.~(\ref{3a}) and (\ref{3b}), the quantities $[ij,kl]$ refer to the  situation where centers of mass displacements of the unit cell contribute to homogenous crystal
strains, while $(ik,jl)$ account for relative displacements of ions within the unit cells, also called internal strains \cite{17} or inner displacements \cite{19}.

There are two independent elastic constants in the case of monolayer hexagonal crystals. They have the dimension of surface tension coefficient.  
Using Voigt's notation ($xx$=1, $yy$=2), one has \cite{10}
 
 \begin{equation}\label{5a}
\mathrm{c}_{11}=\mathrm{C}_{11,11}=[11,11]+(11,11),
\end{equation}
 
 \begin{equation}\label{5b}
\mathrm{c}_{12}=\mathrm{C}_{11,22}=[11,11]-2[11,22]+(11,22),
\end{equation}
 with $\mathrm{c}_{11}$-$\mathrm{c}_{12}$=2$c_{66}$. 

In ab-initio calculations of the elastic constants one distinguishes clamped ion and  relaxed ion terms which we denote with subscripts $ci$ and $ri$, respectively. With the foregoing comments about the physical origin of the square brackets and the round bracket terms, we identify the square brackets terms  on the right hand side of Eqs.~(\ref{5a}) and (\ref{5b}) with the clamped ion contributions while  the sum of square and round brackets terms corresponds to the relaxed ion terms, i.e. the experimentally measured elastic constants. Writing

 \begin{equation}\label{6a}
\mathrm{c}_{11}=\mathrm{c}_{11}\mid_{ri}=\mathrm{c}_{11}\mid_{ci}+\mathrm{c}_{11}\mid_{is},
\end{equation}
 where the subscript $is$ stands for internal strains, we identify
 
\begin{equation}\label{6b}
\mathrm{c}_{11}\mid_{ci}=[11,11]
\end{equation}

\begin{equation}\label{6c}
\mathrm{c}_{11}\mid_{is}=(11,11)
\end{equation}
 
 
 \subsection{Piezoelectric constants}
Born's long wave method \cite{17,19} allows one to calculate the piezoelectric moduli in ionic crystals that do not possess a center of inversion symmetry.  
Within lattice dynamical theory \cite{19} we write the internal strain contribution (suffix $is$) to the piezoelectric stress tensor in 2D as

\begin{equation}\label{7}
e_{i,jl}\mid_{is}=\frac{1}{A_{2D}}\sum_{\kappa\kappa'\kappa''}\sum_{hk}\sqrt{m_{\kappa}}\mathrm{D}^{\kappa\kappa'(1)}_{jh,l}
\Gamma^{\kappa'\kappa''}_{hk}
\frac{Z_{ik}^{(0)\kappa''}}{\sqrt{m_{\kappa''}}}.
\end{equation}

Here $Z_{ik}^{(0)\kappa''}$ is the Fourier transform of the transverse effective charge tensor\cite{19} taken at the $\Gamma$ point of the 2D BZ, 
the other quantities have the same meaning as in the case of Eq.~(\ref{3b}).
Obviously, the right hand side of Eq.~(\ref{7})  takes only into account the inner displacements (strains) of the 
ions. It is therefore also called ionic contribution to the piezoelectric modulus\cite{abinitio}, in contradistinction  to the clamped ion or electronic contribution (see below).
In the case of a 2D hexagonal crystal with $D_{3h}$ symmetry there exists only one independent nonzero piezoelectric stress constant $e_{1,11}$. The internal  strain contribution
$e_{1,11}\mid_{is}$=[1,11] to $e_{1,11}$ is evaluated within analytic theory by means of Eq.~(\ref{7}). In addition to the internal displacement term, the piezoelectric tensor is made up of a second contribution due to the redistribution of the electronic charge cloud upon application of a homogeneous macroscopic strain\cite{martin,baroni,19}.
In analogy with Eq.~(\ref{6a}) we then write
 
 \begin{equation}\label{8b}
e_{1,11}=e_{1,11}\mid_{ri}=e_{1,11}\mid_{ci}+e_{1,11}\mid_{is}.
\end{equation}
 
In $ab$-initio calculations one then distinguishes again relaxed ion and clamped ion contributions. The former should be compared with experimentally measured quantity, the latter is obtained from calculations in absence of internal ionic displacements. While the inner displacements always lead to a reduction of the elastic constants in comparison with the clamped ion contribution, this is not necessarily so for the piezoelectric constants, as we will show in Sect.~\ref{sect3}.


In first-principles calculations the elastic stiffness tensor and piezoelectric tensor coefficients, $e_{ijk}$, are obtained by using density-functional perturbation theory (DFPT)\cite{baroni} as implemented in the Vienna Ab initio Simulation Package (VASP) code\cite{vasp1,vasp2,vasp3,vasp4}. Here, a highly dense $k$-point mesh, 36$\times$36$\times$1, is used to accurately predict these tensor components. The clamped-ion elastic and piezoelectric coefficients are obtained from the purely electronic contribution and the relaxed-ion coefficients are obtained from the sum of ionic and electronic contributions.  Within DFPT the VASP code gives electronic and ionic contribution to the piezoelectric tensor directly. A different approach, also implemented in the VASP code, is based  on the Berry phase concept\cite{abinitio2,berry2}. Here one calculates the polarization for a particular strain, the piezoelectric tensor then follows by calculating the change in polarization due to a strain change. We have used the Berry phase approach with applied uniform strain. 
%
%
At this point, in order to apply strain in a desired direction, the hexagonal primitive cell structure of each material is transformed to a tetragonal one composed of two hexagonal primitive cells\cite{12,16}. A 24$\times$24$\times$1 $k$-point mesh is used to calculate the change in polarization. For all the calculations, the exchange-correlation interactions are treated using the generalized gradient approximation (GGA) within the Perdew-Burke-Ernzerhof (PBE) formulation\cite{cem3} . The single electron wave functions are expanded in plane waves with a kinetic energy cutoff of 600 eV. For the structure optimizations, the Brillouin-zone integrations are performed using a $\Gamma$-centered regular 26$\times$26$\times$1  $k$-point mesh within the Monkhorst-Pack scheme\cite{cem4}. The convergence criterion for electronic and ionic relaxations are set as 10$^{-7}$ and 10$^{-3}$ eV/{\AA} , respectively. In order to minimize the periodic interaction along the $z$-direction the vacuum space between the layers is taken at least 15 \r{A}.

\begin{figure}[ht!]
\includegraphics[scale=0.3]{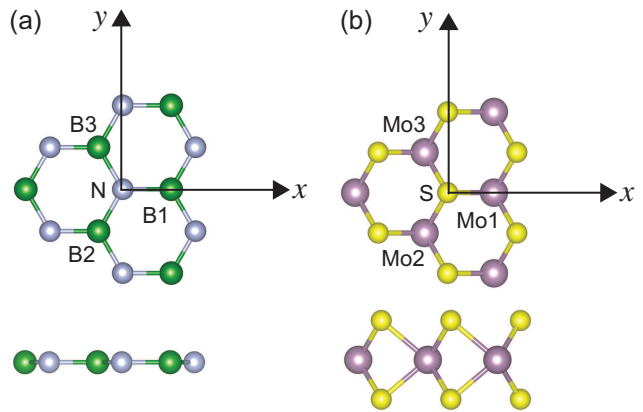}
\caption{\label{fig1} (color online) Schematic plot of (a) a nitrogen (N) atom surrounded by nearest neighbor boron (B) atoms  in 2D-h-BN and (b) a sulfur (S) atom surrounded by nearest neighbor molybdenum (Mo) atoms  in 2H-MoS$_2$. }
\end{figure}

 \section{Composite 2D materials}\label{sect3}
 Two-dimensional hexagonal boron nitride (Fig.~\ref{fig1}(a)), the structurally most simple non centrosymmetric dielectric crystal with point group D$_{3h}$ and two atoms per unit cell, has been used as a prototype for studies of piezoelectricity by ab-initio methods\cite{6,9}. Lattice dynamical theory has been applied to study elastic and piezoelectric effects in single and multilayered crystals\cite{10,11}. Although  h-BN was considered as a specific example, the analytical results are general and are readily applied to materials with the same crystal symmetry such as MoS$_2$ and other TMDCs and TMDOs. 
 
\begin{figure}[ht!]
\includegraphics[scale=0.3]{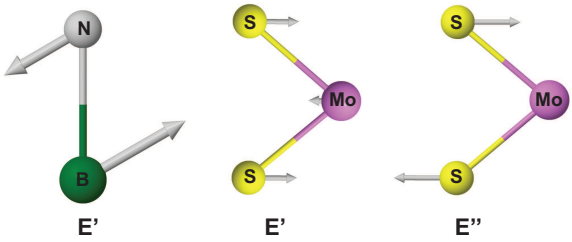}
\caption{\label{fig2} (color online) Atomic displacements of E' mode in 2D-h-BN, and  E' and E'' mode in 2H-MoS$_2$.}
\end{figure}

 Since there are only two atoms per unit cell, in case of B and N, the x$\equiv$1 component of the eigenvector of the optical mode of symmetry E$'$ (Fig.~\ref{fig2}) is given by

\begin{equation}
\xi_1^{\mathrm{E}'}=(\sqrt{\frac{\mu}{m_{\mathrm{N}}}}, -\sqrt{\frac{\mu}{m_{\mathrm{B}}}}),
\end{equation}
where $\mu$=$m_B$$m_N$/M is the reduced mass, with M=$m_B$+$m_N$ the total mass per unit cell. The corresponding eigenfrequency is the doubly degenerate mode $\omega_{E'}$$\equiv$$\omega_{TO}$=$\omega_{LO}$ at $\Gamma$.


The internal strain contribution to $c_{11}$ follows from Eq.~(\ref{3b}) and  is given by\cite{10}
\begin{equation}\label{10}
\mathrm{c}_{11}\mid_{is}\equiv(11,11)_{2D}=-\rho_{2D}\frac{1}{(\omega_{\mathrm{E}'})^2}(\mathrm{D}^{\mathrm{N}\mathrm{B} (1)}_{11,1})^2,
\end{equation} 
where $\rho_{2D}$=M/$A_{2D}$ is the surface mass density. The unit cell area is $A_{2D}$=$a^2$$\sqrt{3}$/2, where $a$=$\mid\overrightarrow{a_1}\mid$=$\mid\overrightarrow{a_2}\mid$ is the length of the in-plane basic vectors of the hexagonal lattice. Likewise the internal strain contribution to the piezoelectric stress constant $e_{1,11}$ follows from Eq. (\ref{7}) and is given by

\begin{equation}\label{9}
e_{1,11}\mid_{is}\equiv[1,11]=\rho_{2D}\mathrm{D}^{\mathrm{N}\mathrm{B} (1)}_{11,1}\frac{1}{(\omega_{\mathrm{E}'})^2}\frac{e^{\ast}_{\mathrm{B}}}{\sqrt{m_{\mathrm{B}}m_{\mathrm{N}}}}.
\end{equation}   

Here the quantity  $e^{\ast}_{\mathrm{B}}$, to be called effective charge of boron ion, stands for the effective charge tensor component $Z_{11}^{(0)\mathrm{B}}$. In the following we will use the notation $e^{\ast}_{\kappa}$ for
$Z_{11}^{(0)\kappa}$. 

We now extend these results to transition metal dichalcogenides and dioxides MX$_2$ where M=Mo, W, Cr and X=O, S, Se, Te. These materials have the same point group symmetry as 2D h-BN. 
 
 As a specific example we consider MoS$_2$, see Fig.~\ref{fig1}(b).  We recall that the nomenclature 2H refers to the most abundant polytype with trigonal prismatic
 coordination between an Mo center surrounded by six sulfide ligands. Each sulfur center is pyramidal and connected to three Mo centers\cite{Schonfeld}, Fig.~\ref{fig1}(b)  
 The internal displacements  that contribute to $c_{11}$ and $e_{1,11}$ are of E$'$ symmetry  where the two S ions on top of each other in each unit cell move in unison and Mo in opposite direction\cite{molina} (Fig.~\ref{fig2}). We then can treat them as an effective particle with mass 2$m_\mathrm{S}$ and charge 2$e^{\ast}_{\mathrm{S}}$=-$e^{\ast}_{\mathrm{Mo}}$. The $x\equiv$1 component of the optical eigenvectors of E$'$ symmetry reads

\begin{equation}\label{17new}
\xi_1^{\mathrm{E}'}=(\sqrt{\frac{\mu}{4m_{\mathrm{S}}}}, \sqrt{\frac{\mu}{4m_{\mathrm{S}}}}, -\sqrt{\frac{\mu}{m_{\mathrm{Mo}}}}),
\end{equation} 
where $\mu$=2$m_{\mathrm{S}}$$m_{\mathrm{Mo}}$/(2$m_{\mathrm{S}}$+$m_{\mathrm{Mo}}$) is the reduced mass. It can be shown that the E'' mode, where the two S ions move
in opposite direction while Mo stays at rest\cite{molina}, does not contribute to $c_{11}\mid_{is}$ or to $e_{1,11}\mid_{is}$ . Writing $\mathrm{D}^{\mathrm{Mo}\mathrm{S} (1)}_{11,1}$ for the matrix element $\mathrm{D}^{\kappa\kappa' (1)}_{11,1}$ where $\kappa$=Mo and $\kappa'$=S, we obtain by means of Eqs.~(\ref{3b}), (\ref{4}) and (\ref{17new})

\begin{equation}\label{11}
\mathrm{c}_{11}\mid_{is}\equiv(11,11)_{2D}=-2\rho_{2D}\frac{1}{(\omega_{\mathrm{E}'})^2}(\mathrm{D}^{\mathrm{S}\mathrm{Mo} (1)}_{11,1})^2,
\end{equation} 
where $\rho_{2D}$=M/$A_{2D}$ is the surface mass density, M=2$m_S+$$m_{Mo}$ the total mass per unit cell, and $A_{2D}$ the basis area of the hexagonal unit cell. 
Similarly we find by means of Eqs.~(\ref{4}), (\ref{7}) and (\ref{17new})

\begin{equation}\label{12}
e_{1,11}\mid_{is}=[1,11]=\sqrt{2}\rho_{2D}\frac{1}{(\omega_{\mathrm{E}'})^2}\mathrm{D}^{\mathrm{S}\mathrm{Mo} (1)}_{11,1}
\frac{ e^{\ast}_{\mathrm{Mo}}}{\sqrt{2m_{\mathrm{S}}m_{\mathrm{Mo}}}}.
\end{equation} 

The translation of Eqs.~(\ref{11}) and (\ref{12}) to other MX$_2$ compounds is straightforward. Using Eqs.~(\ref{6a}) and (\ref{11}) we can write

\begin{equation}\label{13}
\mathrm{c}_{11}\mid_{ri}-\mathrm{c}_{11}\mid_{ci}=-2\rho_{2D}\frac{1}{(\omega_{\mathrm{E}'})^2}(\mathrm{D}^{\mathrm{X}\mathrm{M} (1)}_{11,1})^2,
\end{equation} 
and similarly, combining Eqs.~(\ref{8b}) and (\ref{12}) we get
\begin{equation}\label{14}
e_{1,11}\mid_{ri}-e_{1,11}\mid_{ci}=\sqrt{2}\rho_{2D}\frac{1}{(\omega_{\mathrm{E}'})^2}\mathrm{D}^{\mathrm{X}\mathrm{M} (1)}_{11,1}\frac{ e^{\ast}_{\mathrm{M}}}{\sqrt{2m_{\mathrm{X}}m_{\mathrm{M}}}}.
\end{equation} 
Here $\rho_{2D}$ and $\omega_{\mathrm{E}'}^{(0)}$ are the surface mass density and E' the  optical mode frequency specific for the given MX$_2$ compound. 
In the next section we will use Eqs.~(\ref{13}) and (\ref{14}) for a numerical evaluation of the internal strain coupling $\mathrm{D}^{\mathrm{X}\mathrm{M} (1)}_{11,1}$ and of the effective charge $e^{\ast}_{\mathrm{M}}$.

\begin{table*}
\begin{center}
\caption{\label{table1} $\mathrm{c}_{11}$ in units of N/m,  $e_{1,11}$ in units of 10$^{-10}$ C/m. Inner strain values ($is$) follow from differences 
of relaxed-ion ($ri$) and clamped-ion ($ci$) values, superscripts (1) and (2) refer to input values taken from Refs.\onlinecite{12} and \onlinecite{16}, respectively. 
$\omega_{\mathrm{E}'}$ in units of cm$^{-1}$.}
\begin{tabular}{cccccccc}
\hline \hline 
  Compound   & $\mathrm{c}_{11}\mid_{ri}$  &  $\mathrm{c}_{11}\mid_{ci}$  & $\mathrm{c}_{11}\mid_{is}$  & $e_{1,11}\mid_{ri}$    & $e_{1,11}\mid_{ci}$  & $e_{1,11}\mid_{is}$ & $\omega_{\mathrm{E}'}$    \\ 
\hline
2D h-BN         & 291                          & 300                          & -9                       & 1.38                          & 3.71                           &-2.33                                    &1310                                             \\
2H-MoS$_2$  & 130                          & 153                        & -23                      & 3.64                          & 3.06                            &0.58$^{(\mathrm{1})}$                        &390                                              \\
                      & 133                         & 157                           & -24                    & 4.93                           & 3.20                            & 1.73$^{(\mathrm{2})}$                        &                                                    \\
2H-MoSe$_2$  & 108                        & 131                          & -23                      & 3.92                            & 2.80                         &1.12$^{(\mathrm{1})}$                        &278                                             \\
                         & 107                        & 133                          & -26                      & 5.26                            & 2.92                         & 2.34$^{(\mathrm{2})}$                        &                                                    \\
2H-MoTe$_2$  & 80                        & 101                          & -21                        & 5.43                            & 2.98                         &2.45$^{(\mathrm{1})}$                        &231                                             \\
                         & 84                        & 106                          & -22                        & 6.55                            & 2.75                         & 3.80$^{(\mathrm{2})}$                       &                                                    \\
2H-WS$_2$     & 144                        & 170                          & -26                        & 2.47                            & 2.20                       &0.27$^{(\mathrm{1})}$                        &348                                            \\
                         & 146                        & 175                          & -29                        & 3.76                            & 2.33                       & 1.43$^{(\mathrm{2})}$                       &                                                    \\
2H-WSe$_2$    & 119                        & 147                          & -28                        & 2.71                            & 1.93                       &0.78$^{(\mathrm{1})}$                       &240                                            \\
                         & 120                        & 147                          & -27                        & 3.98                            & 2.05                       & 1.93$^{(\mathrm{2})}$                       &                                                    \\
2H-WTe$_2$    & 89                       & 116                          & -27                         &3.40                           & 1.60                           &1.80$^{(\mathrm{1})}$                          &192                                           \\
                         & 89                        & 115                          & -26                        &5.02                           & 1.75                           & 3.27$^{(\mathrm{2})}$                        &                                                    \\
\hline \hline 
\end{tabular} 
\end{center}
\end{table*}

\section{Numerical results}\label{sect4}

In the previous section we have derived from lattice dynamical theory analytical expressions for the strain contributions $c_{11}\mid_{is}$ and $e_{1,11}\mid_{is}$ to the elastic and piezoelectric constants in 2D hexagonal crystals. On the other hand numerical values of the internal strain quantities can be obtained by taking the differences of the corresponding relaxed-ion and clamped-ion quantities that are calculated by $ab$-$initio$ methods. Whenever the optical phonon frequency $\omega_{\mathrm{E}'}$ is known from $ab$-$initio$ or from experiment, Eqs.~(\ref{13}) and (\ref{14}) allow us to determine the value of the opto-acoustic coupling $\mathrm{D}^{\mathrm{X}\mathrm{M} (1)}_{11,1}$ and of the effective charge $e^{\ast}_{\mathrm{M}}$ respectively for a given material. Such an analysis will also allow us to understand the magnitude of macroscopic quantities such as $c_{11}$ and $e_{1,11}$ on the basis of atomistic concepts that are specific for various materials. The results for 2D h-BN and a series of TMDCs and TMDOs will be presented in Tables \ref{table1}, \ref{table2} and \ref{table3}.  

We first compare the elastic and piezoelectric properties of 2D h-BN with those of MoS$_2$, where the last material is also representative for the other TMDCs.
In Table \ref{table1} we have quoted values of
$c_{11}$ and $e_{1,11}$ calculated  by ab-initio calculations\cite{12,16}  under clamped-ion and relaxed-ion conditions. 
The piezoelectric coefficients $e_{1,11}\mid_{ri}$ and $e_{1,11}\mid_{ci}$ have been obtained\cite{12,16} by the Berry phase method. 
The corresponding values of the internal strains contributions $c_{11}\mid_{is}$ and  $e_{1,11}\mid_{is}$ are then obtained by means of Eqs. (\ref{6a}) and (\ref{8b}). 
Notice that the relative large difference between the values of $e_{1,11}\mid_{ri}$ obtained in Refs.\onlinecite{12} and \onlinecite{16} respectively, leads
to relatively larger difference in the values of $e_{1,11}\mid_{is}$. As mentioned in earlier work\cite{16} we attribute the difference  in the values reported in Ref.\onlinecite{12} to be likely
due to the use of different pseudopotentials and other computational parameters. 
%
%

As a general consequence of Eq.~(\ref{3b}) the internal strains, irrespective of the material, yield a negative contribution to the elastic constants. In the present case, see expressions
of Eq.~(\ref{10}) and (\ref{11}) of $c_{11}\mid_{is}$ for 2D h-BN and 2H-MoS$_2$ respectively and results in Table \ref{table1}.  On the other hand the situation is different for the contribution of internal strains to
$e_{1,11}\mid_{is}$. As Eq.~({\ref{7}) suggests, the sign of the effective ionic charges determines the sign of  $e_{1,11}\mid_{is}$. While the clamped-ion piezoelectric constant $e_{1,11}\mid_{ci}$ is larger for 2D h-BN than for 2H-MoS$_2$, the relaxed-ion piezoelectric modulus $e_{1,11}\mid_{ri}$, which corresponds to the experimentally measured quantity \cite{14,15}, is larger for 2H-MoS$_2$ than for 2D h-BN (see Table~\ref{table1}). In the latter material the internal strain component $e_{1,11}\mid_{is}$ leads to a reduction of $e_{1,11}$ and in the former to an increase.  Notice that the large absolute value of $e_{1,11}\mid_{is}$
 of 2D h-BN is in accordance with the large ionic contribution to the static dielectric response obtained from ab-initio finite electric field calculations in BN nanotubes \cite{nw1}. 
A positive value of $e_{1,11}\mid_{ci}$ has also been obtained by analytical  calculations \cite{Droth} using the Berry phase method. 
We observe that the opposite sign of the ionic contribution $e_{1,11}\mid_{is}$ in comparison with the electronic contribution $e_{1,11}\mid_{ci}$ as is here the case for 2D h-BN is not uncommon in other piezoelectric materials and has been found also in 3D III-IV semiconductors.\cite{abinitio}

%

\begin{table}[ht]
\begin{center}
\caption{\label{table2} Surface mass density $\rho_{\mathrm{2D}}$ (units 10$^{-6}$ kg/m$^2$); optical mode frequency $\omega_{\mathrm{E}'}$  (units cm$^{-1}$); 
Opto-acoustic coupling $\mathrm{D}^{\mathrm{\cdot\cdot} (1)}_{11,1}$ (units 10$^{19}$ cm s$^{-2}$); effective charges $e^{*}_{\mathrm{M}}$ of metal ions (units elementary charge $e$= 1.602 $\times$10$^{-19}$ C), (a) present theory, superscripts  (1) and (2) refer to $e_{1,11}\mid_{is}^{(1)}$ and $e_{1,11}\mid_{is}^{(2)}$ from Table \ref{table1}, respectively, 
(b) effective charge calculated with DFPT.}
\begin{tabular}{cccccc}
\hline \hline 
  Compound       &    $\rho_{\mathrm{2D}}$   & $\omega_{\mathrm{E}'}$ & $\mathrm{D}^{\mathrm{\cdot\cdot} (1)}_{11,1}$   & $e^{\ast (\mathrm{a})}_{\mathrm{M}}$    & $e^{\ast (\mathrm{b})}_{\mathrm{M}}$   \\ 
\hline
2D h-BN            & 0.76                   & 1310                                & -8.44                                                                    & +2.78                                                        & 2.71                          \\
2H-MoS$_2$    & 3.14                 &390                                   & -1.99                                                                    & -0.41$^{(\mathrm{1})}$                            & -1.06                                 \\
                         &                        &                                         &                                                                              & -1.21$^{(\mathrm{2})}$                            &                                    \\
2H-MoSe$_2$ & 4.42                 &278                                  & -0.91                                                                    & -0.70$^{(\mathrm{1})}$                            & -1.84                                 \\    
                         &                        &                                         &                                                                             & -1.45$^{(\mathrm{2})}$                            &                                    \\                                                  
2H-MoTe$_2$ & 5.34                 &231                                  & -0.63                                                                    & -1.59$^{(\mathrm{1})}$                            & -3.28                                 \\    
                        &                        &                                         &                                                                             & -2.46$^{(\mathrm{2})}$                            &                                    \\                         
2H-WS$_2$   & 4.70                &348                                  & -1.14                                                                      & -0.17$^{(\mathrm{1})}$                            & -0.53                                 \\    
                         &                        &                                         &                                                                            & -0.92$^{(\mathrm{2})}$                            &                                    \\                         
2H-WSe$_2$  & 5.60                &240                                  & -0.68                                                                      & -0.49$^{(\mathrm{1})}$                            & -1.22                                 \\    
                         &                        &                                         &                                                                            & -1.21$^{(\mathrm{2})}$                            &                                    \\
2H-WTe$_2$ & 6.60                &192                                  & -0.51                                                                      & -1.11$^{(\mathrm{1})}$                            & -2.6                                 \\    
                        &                        &                                         &                                                                             & -2.01$^{(\mathrm{2})}$                            &                                    \\                                     
       \hline \hline 
\end{tabular} 
\end{center}
\end{table}

We now show that this different behavior of 2D h-BN and 2H-MoS$_2$ is due to the opposite sign of the effective charges $e^{\ast}_{\mathrm{B}}$ and $e^{\ast}_{\mathrm{Mo}}$. 
We first calculate the effective charges by a semi-analytical method, inverting Eq. (\ref{7}). Thereby we first  determine the opto-acoustic coupling  $\mathrm{D}^{(1)}$ from $ab$-initio values
of the elastic constants. In case of 2D h-BN we start from Eq. (\ref{10}), insert the numerical values for $\mathrm{c}_{11}\mid_{is}$ (Table \ref{table1}), the optical mode frequency $\omega_{\mathrm{E}'}$=1310 cm$^{-1}$, the surface mass density $\rho_{2D}$=7.59$\times$10$^{-8}$ g/cm$^{2}$, and obtain the value for $\mathrm{D}^{\mathrm{NB} (1)}_{11,1}$ quoted in Table \ref{table2}. Here we have retained the negative value of $\mathrm{D}^{\mathrm{NB} (1)}_{11,1}$ which is consistent with force constant model calculations of the dynamical matrix\cite{10}. Notice that we have taken into account that the direction of the $x$-axis in the present paper is different from Refs.\cite{10,11}. Likewise we proceed for 2H-MoS$_2$ starting from Eq. (\ref{11}), where we have used $c_{11}\mid_{is}$ from Table \ref{table1} and $\omega_{\mathrm{E}'}$=390 cm$^{-1}$ and $\rho_{2D}$=3.14$\times$10$^{-7}$ g/cm$^{2}$. See $\mathrm{D}^{\mathrm{..} (1)}_{11,1}$ in Table \ref{table2}.  The large difference in absolute value between $\mathrm{D}^{\mathrm{NB} (1)}_{11,1}$ and $\mathrm{D}^{\mathrm{SMo} (1)}_{11,1}$ is due to the fact that the interatomic forces in 2D  h-BN are considerably stronger than in 2H-MoS$_2$, in accordance with the different phonon spectra for h-BN\cite{10,serrano} and MoS$_2$\cite{molina,waka}. 
 We next turn to Eqs. (\ref{9}) and (\ref{12}), insert the corresponding values for $e_{1,11}\mid_{is}$ from Table \ref{table1} as well as $\mathrm{D}^{\mathrm{\cdot\cdot} (1)}_{11,1}$ from Table \ref{table2}, the corresponding frequencies $\omega_{\mathrm{E}'}$, masses and densities, and solve with respect to $e^{\ast}_{\mathrm{M}}$. The results are shown for $e^{\ast}_{\mathrm{B}}$ and $e^{\ast}_{\mathrm{Mo}}$ in column 5 of Table \ref{table2}. In column 6 we have quoted values of $e^{\ast}_{\mathrm{M}}$ obtained directly by DFPT calculations. Although the magnitude of $e^{*}_{\mathrm{Mo}}$ depends strongly on the values of the input quantity  $e_{1,11}$$\mid_{ri}$ (see Table  \ref{table1}), we obtain $e^{*}_{\mathrm{Mo}}$ $<$ 0 and $e^{*}_{\mathrm{B}}$  $>$ 0, in agreement with results from ab-initio calculations \cite{25}.  We attribute the positive value of $e^{*}_{\mathrm{B}}$ to the larger electronegativity 3.0 of N in comparison with 2.0 of B (see Ref.\onlinecite{atkins}), notwithstanding  that there is an opposite electron transfer from N to B within the $\pi$ bond.  In case of MoS$_2$ we may assume that the 4$d^5$5$s^1$ valence electrons of Mo participate in the bonds with the six surrounding S atoms, each of which has valence configuration of 3$s^2$3$p^4$.
Thereby the shielding effect at Mo is decreased and the excess electrons of S lead to an effective negative charge $e^{*}_{\mathrm{Mo}}$. Obviously the large difference between the values -0.41$^{{(1)}}$ and -1.21$^{(2)}$ of 
$e^{*(a)}_{\mathrm{Mo}}$ is due to the different values of $e_{1,11}\mid_{is}$, 0.58 and 1.73 obtained respectively from Refs.\onlinecite{12}  and \onlinecite{16}, see Table ~\ref{table1}.

\begin{table}
\begin{center}
\caption{\label{table3} Symbols have the same meaning and unit as in Table \ref{table2}. }
\begin{tabular}{cccccccc}
\hline \hline 
  Compound   & $\rho_{2\mathrm{D}}$  &  $\omega_{\mathrm{E}'}$   & $\mathrm{D}^{\mathrm{\cdot\cdot} (1)}_{11,1}$   & $e^{*(a)}_{\mathrm{M}}$  & $e^{*(b)}_{\mathrm{M}}$   \\  
  \hline
CrS$_2$    & 2.41  & 415 & -1.44 & -1.79 & -2.44 \\ 
CrSe$_2$  & 3.10  & 317  & -0.82 & -2.33 & -3.06  \\ 
CrTe$_2$  & 4.87   & 262  &-0.58  & -3.19 & -4.13  \\ 
CrO$_2$  &  2.34   &591   &-2.01   &-0.189 & 1.334 \\
MoO$_2$ & 3.06    & 522  & -1.91  &0.215   & 2.875 \\
WO$_2$ & 5.15      & 491   & -1.44  &0.382  &3.326 \\
 \hline \hline 
\end{tabular} 
\end{center}
\end{table}


In Table \ref{table3} we present results of similar calculations for Cr based dichalcogenides and some TMDOs. Although these materials have not been synthesized  so far, their mechanical and dynamical stability have already been shown by first principles calculations\cite{mx2-deniz}.

From Table \ref{table2} and \ref{table3} if follows that for all TMDCs, notwithstanding quantitative differences, the effective charges $e^{*(a)}_{\mathrm{M}}$ and $e^{*(b)}_{\mathrm{M}}$ obtained by the present analytical calculations or by DFPT respectively, are negative. Notice also that the absolute values increase in the order X=S, Se, Te, in accordance with the charge number Z of these elements, see Table \ref{table4}. For the same chalcogene, $e^{*}_{\mathrm{M}}$ decreases in absolute value in the order M= Cr, Mo, W, and we conclude that the decrease of shielding mentioned before for Mo is less (more) efficient for heavier (lighter)  metals. Our results are in quantitative agreement with recent Born effective charge tensor calculations for MoS$_2$, MoSe$_2$, WS$_2$ and WSe$_2$ \cite{nw2}. 

For TMDOs the $e^{*}_{\mathrm{M}}$ values obtained by analytical theory and direct ab-initio calculations differ by an order of magnitude, also the values of $e^{*}_{\mathrm{M}}$ obtained by $ab$-$initio$ are all positive. Obviously, the large electronegativity 3.4 \cite{atkins} of O leads to a negative internal strain  contribution to $e_{1,11}$ as is the case for 2D h-BN. The large quantitative difference between analytical and $ab$-$initio$ calculations suggests that the concept of "effective" ionic charge\cite{w} used in the analytical  theory gives fair results for TMDCs but breaks down for TMDOs. In the latter case the deformability of the electron cloud\cite{19} of the metal ions upon inner displacements should not be neglected while in Eqs. (\ref{9}), (\ref{12}) and (\ref{14}) we have assumed a rigid ion model. 

We observe that the opto-acoustic coupling $\mathrm{D}^{\mathrm{XM(1)}}_{11,1}$ given in column 4 of Tables \ref{table2} and \ref{table3} are a measure of the X-M bond strength. 
We see that for a given metal M this quantity decreases in absolute value with increasing size (i.e. charge number Z) of the X atom, likewise for a given X, $\mathrm{D}^{\mathrm{XM(1)}}_{11,1}$ decreases with increasing size of the metal atom, see Table \ref{table5}. 


We have also carried out calculations of $e_{1,11}\mid_{ir}$ and $e_{1,11}\mid_{ci}$ by using density functional perturbation theory  instead of the Berry phase method and then determined  $e^{*}_{\mathrm{M}}$.  The results are quoted in Table ~\ref{table6}. Although the overall agreement with directly calculated values $e^{*}_{\mathrm{M}}$ is less satisfactory than in the case of Tables \ref{table2} and \ref{table3}, the material dependent trends are similar.  


\begin{table}[ht]
\begin{center}
\caption{\label{table4} Trends in effective ionic charges $e^{*}_{\mathrm{M}}$ (units elementary charge $e$) for 2H-MX$_2$; integers 16, 24 ,... are charge numbers Z.}
\begin{tabular}{cccc}
\hline \hline 
     & Cr$^{24}$  & Mo$^{42}$  & W$^{74}$  \\ 
\hline
  S$^{16}$   &  -2.44  &  -1.06   &      -0.53                \\
\hline
   Se$^{34}$  & -3.05   &  -1.84 &   -1.22     \\
\hline
 Te$^{52}$   & -4.19    &  -3.28  &  -2.60            \\
       \hline \hline 
\end{tabular} 
\end{center}
\end{table}

\begin{table}[ht]
\begin{center}
\caption{\label{table5} Trends in opto-acoustic  couplings $\mathrm{D}^{\mathrm{\cdot\cdot} (1)}_{11,1}$ in units 10$^{19}$ cm s$^{-2}$. }
\begin{tabular}{cccc}
\hline \hline 
     & Cr$^{24}$  & Mo$^{42}$  & W$^{74}$  \\ 
\hline
  S$^{16}$   &  -1.44  &  -1.43   &      -1.04               \\
\hline
   Se$^{34}$  & -0.82   &   -0.91 &   -0.68     \\
\hline
 Te$^{52}$   & -0.58    &  -0.63  &  -0.51            \\
       \hline \hline 
\end{tabular} 
\end{center}
\end{table}


\begin{table}
\begin{center}
\caption{\label{table6} Opto-acoustic coupling $\mathrm{D}^{\mathrm{\cdot\cdot} (1)}_{11,1}$ in units 10$^{19}$ cm s$^{-2}$;  $\omega_{\mathrm{E}'}$ in units of cm$^{-1}$; $e_{1,11}\mid_{is}$ (units 10$^{-10}$ C/m); surface mass density $\rho_{2D}$ (units of 10$^{-6}$ kg/m$^{2}$); effective charges $e^{*}_{\mathrm{M}}$ of metal ions (units $e$), (a) present theory, and (b) ab-initio calculations using density functional perturbation theory.}
\begin{tabular}{ccccccc}
\hline \hline 
  Compound   & $\mathrm{D}^{\mathrm{\cdot\cdot} (1)}_{11,1}$   & $\omega_{\mathrm{E}'}$ & $e_{1,11}$$\mid_{is}$ & $\rho_{2\mathrm{D}}$  & $e^{*(a)}_{\mathrm{M}}$  & $e^{*(b)}_{\mathrm{M}}$   \\  
  \hline
CrS$_2$   & -1.440 &415   & 1.30   & 2.408  & -0.971&-2.443\\ 
CrSe$_2$  & -0.817 & 317  &  1.80  & 3.100 & -1.329 & -3.056\\ 
CrTe$_2$  & -0.582 &  262 &   2.68 &  4.872 &-1.950&-4.130\\ 
MoS$_2$  & -1.429 &  377 &   0.56 &  3.029 & -0.376&-1.063\\ 
MoSe$_2$ & -0.906 &  278 &   1.09 &  4.421 &-0.677 &-1.840\\ 
MoTe$_2$ & -0.626 &  231 &  2.16 &  5.344 & -1.396 &-3.277\\ 
WS$_2$  & -1.135 &  348 &   0.28 &  4.695 & -0.179&-0.526\\ 
WSe$_2$ & -0.681 &  240 &  0.73  &  5.600 & -0.460& -1.224\\ 
WTe$_2$ & -0.508 &  192 &  1.74 &  6.671 & -1.071&-2.600\\ 
CrO$_2$ & -2.010 &  591 &   -0.38 &  2.335 & 0.300& 1.334\\
MoO$_2$ & -1.910 &  522 &   -0.98 &  3.063 & 0.658& 2.875\\
WO$_2$ & -1.440 &  491 &   -1.15 &  5.153 & 0.745& 3.326\\
 \hline \hline 
\end{tabular} 
\end{center}
\end{table}


\section{Conclusion}
We have presented a synthesis  of lattice dynamical theory and $ab$-$initio$ calculations results for the description of elastic and piezoelectric properties of
two dimensional ionic crystals with hexagonal lattice structure. As specific examples we investigated 2D h-BN as well as 2H-TMDCs and 2H-TMDOs of composition
MX$_2$ where M is a transition metal ion and X a chalcogen or oxygen ion. Such a study allowed us to separate quantitatively electronic and ionic contributions
to the elastic and piezoelectric constants. We have investigated the validity of microscopic concepts such as rigid ion model and effective
ionic charges for various MX$_2$ compounds. 

Further we have been able to discern trends in the values of the opto-acoustic coupling $\mathrm{D}^{\mathrm{XM} (1)}_{11,1}$ and of the optical mode frequency
$\omega_{\mathrm{E}'}$ as function of the atomic composition.  For a given metal M, $\mathrm{D}^{\mathrm{XM} (1)}_{11,1}$ and $\omega_{\mathrm{E}'}$ decrease in absolute
value with increasing atomic number Z of the chalcogen X ion, and similarly for a given X, these quantities decrease with increasing Z of  M. These properties
reflect the strength of the chemical bonds of the M ion to the 6 surrounding X ions. We find that the effective charge $e^{*}_{\mathrm{M}}$ of the metal ion is negative for all
TMDCs while $e^{*}_{\mathrm{B}}$ in 2D h-BN is positive. This difference in sign entails that the inner strain contribution $e_{1,11}$$\mid_{is}$ leads to a reduction of $e_{1,11}$
in case of 2D h-BN and to an increase in case of 2H-MoS$_2$ and other TMDCs. 
{\textit{Acknowledgments}}. \label{agradecimientos} The authors acknowledge useful discussions with L. Wirtz, and A. Molina-Sanchez. This work was supported by the Methusalem program, and the Flemish Science Foundation (FWO-Vl).  Computational resources were provided by HPC infrastructure of the University of Antwerp (CalcUA) a division of the Flemish Supercomputer Center (VSC), which is funded by the Hercules foundation.
%
%

%

\end{document}